# Singular points of polarizations in the momentum space of photonic crystal slabs

Weimin Ye[1*], Yang Gao[4], and Jianlong Liu[2,3†]

1. College of Advanced Interdisciplinary Studies, National University of Defense Technology, Changsha, 410073, China

2. College of Physics, Harbin Institute of Technology, Harbin 150001, China

3. College of Physics and Optoelectronic Engineering, Harbin Engineering University, Harbin 150001, China

4. College of Electronic Engineering, Heilongjiang University, Harbin 150080, China

[*]wmye72@126.com

[†]liujl@hit.edu.cn

**Abstract**

**Bound states in the continuum (BICs), circularly polarized states (*C* points) and degenerate states are all of three types of singular points of polarization in the momentum space. For photonic crystal slabs (PhCSs) with linearly polarized far fields, BICs were found to be the centers of polarization vortices and attracted more attention in the previous studies. Here, we theoretically demonstrate that the far fields of PhCSs can exhibit remarkably diverse polarizations due to the robust existences of *C* points in the continuum. Only a pair of *C* points with identical handedness and opposite topological charge can be annihilated together. Continuously fine tuning of the structure parameters of PhCSs without breaking their symmetry, a pair of *C* points with identical topological charge and opposite handedness are able to merge into a BIC, then the BIC splits into *C* points again. Interestingly, a Dirac-degenerate BIC with one half of topological charge is observed when two pairs of *C* points with identical topological charge from the upper and lower band, respectively, simultaneously merge at the Dirac-degenerate point. The law of topological charge conservation is verified to play an important role in the evolutions and interconversions between different types of polarization singularities. Our findings might shed light on the origin of singular points of polarization, could open a gateway towards the applications of them in the generation and manipulation of vector beams.**

Analogous to polarization singularities in real space of the monochromatic electromagnetic fields (vector fields) [1-4], singular points of polarizations in the momentum space are referred to as eigenmodes of optical systems with undefined polarization directions in the far fields. There are only three types of singular points of polarizations: *V* points (where the intensity of far field vanishes), *C* points (where the far field is pure circular polarization), and degenerate points at band degeneracies (where the far-field polarization is undetermined). Owing to the advantages in the designs, fabrications and on-chip applications, Bloch leaky modes supported by photonic crystal slabs (PhCSs) [5,6] can present all types of singular points of polarizations in their momentum space, which has attracted considerable attentions[7-19]. Worth noting that leaky modes of PhCSs at *V* points are lossless and cannot couple with free-space radiations, they are indeed the optical bound states in the continuum (BICs) with infinite lifetimes [7-11]. For linearly polarized far fields, BICs (*V* points in the momentum space) have been verified to be the momentum-space vortex centers exhibited by the polarization vectors [12-14]. This topological property ensures the robust existence of BICs [12] and opens new applications for BICs in the creation of vector lasers [15] and ultrahigh-Q guided resonances [16]. At *C* points, PhCSs only couple with the circularly polarized free-space radiations with identical handedness. It leads to strong chiroptical effects and can be directly utilized to modulate the ellipticity of light. This phenomenon was verified by employing the PhCSs with a pair of *C* points spawning from an eliminated BIC at Γ point [17]. Besides, optically pumped BIC lasers were realized in the square-lattice PhCSs with double-degenerate BICs at Γ point [18].

In the previous studies, the frequencies of all reported BICs [7-16], *C* points [17] and degenerate points [18,19] in the momentum space of PhCSs were below the diffraction limit. That is, only the zero-order diffraction of far fields was the propagating wave in free space. These singular points of polarizations [7-19] could not arrive at the boundary of the first Brillouin zone (FBZ) before they became guided modes. It is due to the fact that for any leaky mode of PhCSs on the boundary of FBZ, at least the 0th- and 1st (or -1st) -order diffractions of them are free-space radiations. Although the occurrence of more than one order diffractions could bring more restrictions on constructing the BICs, the interferences among them provide a degree of freedom to efficiently modulate the far-field polarization states of PhCSs beyond the diffraction limit. It is, therefore, expected that singular points of polarizations near the boundary of FBZ could exhibit different topological nature from BICs below the diffraction limit [12].

In this Letter, focusing on the polarization singularities near the corner of FBZ, we study Dirac-degenerate points and BICs at K point (K-point BICs) and *C* points supported by the honeycomb-lattice PhCSs [20-22]. They are direct analogue to graphene and could introduce two degrees of freedom, pseudospins, and valleys to the photons, which are widely used in topological photonics [23]. Distinct from the linearly polarized far fields [9,12-16] and *C* points achieved by breaking the in-plane inversion symmetry of the PhCSs[17], we theoretically demonstrate that *C* points can emerge near K-point BICs and Dirac-degenerate points of PhCSs with $180^0$ rotational symmetry around *z*-axis (the in-plane inversion symmetry) and time-reversal symmetries. As a result, far fields of the PhCSs exhibit remarkable polarization diversity. Circular polarizations, linear and elliptical polarizations with variant orientations and ellipticities could appear near K point. Continuously varying the geometric parameters of PhCSs without breaking their spatial symmetry, the appearance and disappearance of K-point BICs (the Dirac-degenerate BICs) are found to be accompanied by merging and generating one (two) pair(s) of *C* points, respectively. Thus, we propose to use the winding numbers of the trajectories of far-field polarization states on Poincaré sphere around the $S_3$ axis to generally define the topological charges carried by the polarization singularities. The conservation law of topological charges carried by all singular points provides an intuitive understanding of the robust existence of *C* points in the continuum (above the light line) and the interconversions of different types of polarization singularities.

Figure 1(a) shows our considered 2D PhCS composed of a honeycomb array (a lattice constant *a*) of cylindrical holes (identical diameter *D*) etched in a free-standing dielectric slab (a thickness *h* and a refractive index *n*=2.02 corresponding to $Si_3N_4$ [9]). Due to the *z*- and *y*-mirror symmetry (invariance under the operation $\sigma_z$ changing *z* to –*z* and $\sigma_y$ changing *y* to –*y*), all eigenmodes of the PhCS could be divided into TM-like (defined by $\sigma_z=-1$) and TE-like ($\sigma_z=1$) modes. The eigenmodes with the *y* component of Bloch wave vector $k_y$ equal to zero could be further divided into even (defined by $\sigma_y=1$ in the Letter) and odd ($\sigma_y=-1$) modes [6]. Figure 1(b) depicts the FBZ of a honeycomb lattice. At its corners, three **K** points, denoted by $\mathbf{K}_j$, *j*=1, 2, 3, are equivalent to each other. For leaky modes at $\mathbf{K}_1$ point, the 0th-order and the two 1st-order diffractions with in-plane wave vectors equal to $\mathbf{K}_2$ and $\mathbf{K}_3$, respectively, are three propagating waves in free space, which provides six leakage channels. Due to $C_{3V}$ symmetry of the PhCSs at **K** points, the coefficients of the three diffractions are not independent. Only one (two) leakage channels superposed by them is (are) symmetry-compatible with the

non-(Dirac-)degenerate leaky modes at **K** points. For example, the electric field of single leakage channel compatible with a non-degenerate odd BIC at **K**$_1$ point is given by $\boldsymbol{E}(\sigma_y = -1) = \boldsymbol{E}_{K_1}^{(TE)} + \boldsymbol{E}_{K_2}^{(TE)} + \boldsymbol{E}_{K_3}^{(TE)}$ ($\boldsymbol{E}_{K_j}^{(TE)}$ denotes the electric-field of TE polarized plane wave with in-plane wave vector equal to **K**$_j$). Thus, none of K-point BICs is symmetry-protected BIC, which is different from Γ-point BICs [12,17]. By tuning geometric parameters, the PhCSs with (*D*, *h*) equal to (0.3*a*, 0.9074*a*) is found to support a non-degenerate TE-like odd BIC at **K**$_1$ point with a real normalized frequency *ωa/2πc* equal to 0.7998, which is located at the bottom of the band [Fig. 1(c)] (for the detailed band structures see Fig. s1 in Supplementary Material Note 1).

To look into the topological natures of the BIC, we examine the distributions of the far-field polarization states of the eigenmodes supported by the PhCSs in the vicinity of **K$_1$** point [Fig. 2(a)] (For details of the method used to obtain the polarization states, see the Supplementary Material Note 2). Compatible with the odd symmetry of the BIC, the far fields of the eigenmodes with *k$_y$* equal to zero are TE polarizations (shown as a short line in the *y* direction). Interestingly, a pair of *C* points with the different handedness shown as the red and blue disc (corresponding to right and left-handed circular polarizations, respectively) appear at (*k$_x$*, *k$_y$*) equal to (-1.022*K*, -0.032*K*) and (-1.022*K*, 0.032*K*), respectively. When the handedness and the in-plane wave vector of the circularly polarized incident light match those of the *C* point, the reflectance spectrum of the PhCS exhibits a Fano line-shape curve with the resonant frequency approximately equal to that of the eigenmode at the *C* point. Utilizing a simplified scattering-matrix model [24] to study the couplings between fields inside the PhCS and the free-space leakages, we prove that both the *C* points and the K-point BIC are achieved by the interferences among multiple radiation channels (for details, see the Supplementary Material Note 3). Note that there is an *L*-line [2-4], where the far fields are linear polarizations, along the boundary between the right- and left-handed ellipses. The predominant component of the electric field on the *L*-line is TM polarization (shown as a short line in the *x* direction). The point of intersection between the *L*-line and the line *k$_y$*=0 in the momentum space is just **K**$_1$ point (for details, see Fig. s4 in Supplementary Material Note 4). Owing to the *y*-mirror symmetry of the PhCS, the polarization direction of the far field at **K**$_1$ point is undefined. Thus, the non-degenerate eigenmode at **K**$_1$ point is a BIC with an infinite quality factor (*Q* value) [12]. The diverse far-field polarization states including linear, circular and elliptical polarizations with different orientations cannot be simply described by a polarization vortex.

Here, we map the varying far-field polarization states on the shell of Poincaré sphere, when Bloch wave vector $\bm{k}$ moves along an anticlockwise closed loop $L$ enclosing singular points of polarizations in the momentum space. For a far-field polarization state with an orientation angle $\psi(\bm{k})$ and ellipticity angle $\chi(\bm{k})$, the longitude $\varphi(\bm{k})$ and the latitude $\theta(\bm{k})$ of its corresponding point on the shell of Poincaré sphere are equal to $2\psi$ and $2\chi$, respectively. Similar to $C$ point located at the north and south poles with undefined longitude or orientation angle, the longitudes of the BIC and degenerate points (not shown on the shell of Poincaré sphere) are undefined. Thus, limited to the closed loop $L$ enclosing only one BIC ($C$ point or degenerate point on one of the degenerate bands), the topological charge $q_B$ ($q_C$ or $q_D$) carried by the BIC ($C$ point or the degenerate point) can be defined as half of the winding number $n_w$ (an integer) of the corresponding closed trajectory $G(L)$ around the $S_3$ axis on the shell of Poincaré sphere. That is,

$$q_{B(C,D)} = \frac{1}{2\pi} \oint_L d\psi = \frac{1}{4\pi} \oint_{G(L)} d\varphi = \frac{n_w}{2}. \qquad (1)$$

Meanwhile, the winding number $n_w$ of the closed trajectory $G$ is double of the sum of topological charges carried by all polarization singularities enclosed within the corresponding closed loop $L$ in the momentum space. The topological charge defined in Eq. (1) is consistent with that of $C$ point and $V$ point in real space [3, 4]. When the far field is linearly polarized, the topological charge is just the winding number of the polarization vortex [12].

Three anticlockwise circles, denoted by $L_1$, $L_2$, $L_3$ in Fig. 2(a), enclosing the BIC with the radii $R_k$ equal to $0.016K$, $0.033K$, $0.056K$ and the centers ($k_x$, $k_y$) equal to (-$K$, 0), (-1.01$K$, 0.015$K$), (-$K$, 0), respectively, are selected to map the far-field polarization states on them to the shell of Poincaré sphere. Figures 2(b)-2(d) show that the winding numbers of the three closed trajectories corresponding to circles $L_1$, $L_2$ and $L_3$ are equal to -2, -1 and 0, respectively. Since the amount of $C$ points enclosed by circles $L_1$, $L_2$ and $L_3$ is 0, 1 and 2, the topological charge [Eq.(1)] carried by the BIC and the $C$ point is -1 and 1/2, respectively. [For the projected trajectories (letting $S_3$=0) of the three-dimensional trajectories in Figs. 2(b)-2(d) on the $S_1$-$S_2$ plane, see Figs. s5 in Supplemental Material Note 5]. The conserved sum of the topological charges including the pair of $C$ points and the BIC in the momentum space shown in Fig. 2(a) is zero, which plays an important role in their generations, evolutions and annihilations.

To see the behaviors of these singularities in the momentum space, we slightly change the diameter $D$ of the cylindrical holes etched in the PhCS [Fig. 1(a)] without breaking its symmetry. Noting that there are usually six (three) independent leakage channels in free space for leaky modes with Bloch wave vectors (along the ΓK direction or on the boundary of FBZ) close to K point, the increased amount of leakage channels destroys the robustness of K-point BICs under small changes of geometric parameters of PhCSs. Without violating the conservation law of topological charges, the K-point BIC [in Fig. 2(a) with a topological charge of -1] should spawn two *C* points. Figure 3(a) exhibits that the K-point BIC just emerges at the point of intersection between the trajectories of a pair of *C* points [denoted by -R1/-L1 in Fig. 3(a) with opposite handedness and identical topological charges of -1/2]. When *D* is relatively small, this pair of *C* points locate inside the FBZ and close to $K_1$ point. With the continuous increment of *D*, the pair of *C* points move outwards and cross each other at $K_1$ point. Here, the BIC with an infinite *Q* value appears [Figs. 3(b)-3(d)]. Meanwhile, another pair of *C* points outside the FBZ [shown in Fig. 2(a) with the topological charge of 1/2 and denoted by +R2/+L2 in Fig. 3(a)] move inwards. These two pairs of *C* points with identical handedness and opposite charges are annihilated with each other when *D* reaches 0.305*a*. The disappearance of polarization singularities is consistent with the sum of the topological charges equal to zero [Fig. 3(a)].

Furthermore, we study the behaviors of Dirac-degenerate BICS. The PhCS shown in Fig. 1(a) with *D*=0.286*a* and *h*=1.036*a* supports a TE-like double-degenerate BIC at the $\mathbf{K}_1$ point with a real normalized frequency of 0.7311, where two TE-like bands of the PhCS linearly cross each other and form the Dirac cone dispersion [Fig. 4(a)] (for the detailed band structures, see Fig. s6 in Supplementary Material). Figure 4(d) displays the far-field polarization states of eigenmodes in the upper band of the Dirac cone. Due to the *y*-mirror symmetry of the PhCS and the undefined polarization states of double-degenerate modes at $\mathbf{K}_1$ point, the far fields of eigenmodes with Bloch wave number $k_y$=0 are TM and TE polarization on the left and right sides of $\mathbf{K}_1$ point, respectively. Based on Eq. (1), we can obtain that the topological charge carried by the Dirac-degenerate BIC is 1/2. Figure 4(e) presents similar results of the eigenmodes in the lower band. Since the Dirac cone dispersion and double-degenerate modes at $\mathbf{K}_1$ point are protected by the spatial symmetry of the honeycomb-lattice PhCSs, when varying the diameter *D* of cylindrical holes etched in the PhCSs, the double-degenerate point is preserved at $\mathbf{K}_1$ point. But, it is not a BIC now. Figures 4(b) and 4(c) [4(f) and 4(g)], respectively, show the far-field polarization states of eigenmodes in the upper and lower band of the Dirac cone when the diameter *D* decreased (increased) to 0.256a (0.306*a*). Here, the topological charge carried by the Dirac-degenerate point at $\mathbf{K}_1$ point is -1/2, which is opposite to the Dirac-degenerate BIC [Figs. 4(d) and 4(e)]. The increased (reduced) topological charges originate from the merged (spawned) two pairs of C points carried the identical topological charges of 1/2 at $\mathbf{K}_1$ point. When $D$ is equal to 0.256$a$, the two pairs of *C* points are outside the FBZ, denoted by +R1/+L1 in the upper band [Fig. 4(b)] and +R2/+L2 in the lower band [Fig 4(c)], respectively. Continuously increasing $D$, the two pairs of *C* points move inwards and cross each other at $K_1$ point into the FBZ. The pair of *C* points denoted by +R1/+L1 (+R2/+L2) enters the lower (upper) band in Fig. 4(g) [Fig. 4(f)] (for details of the distributions of quality factors (Q value) of the eigenmodes, see Fig. s7 in Supplementary Material).

In conclusion, focusing on polarization singularities near the boundary of the first Brillouin zone, we have theoretically demonstrated that the robust existence of *C* point above the light line and the coexistence of different types of singular points of polarizations in the momentum space. Only a pair of *C* points with identical handedness and opposite topological charge can be annihilated together. It leads to diverse polarization states in the far-field radiations of photonic crystal slabs (e.g. circular polarization, elliptical and linear polarizations with variant orientations). Distinct from the reported robust BICs inside the FBZ [12], the emergence of a BIC at the corner of FBZ just results from merging a pair of *C* points with identical topological charge and opposite handedness**.** Continuously varying the geometric parameters of PhCSs without breaking their spatial symmetry, the BIC could split into C points again. Interestingly, the topological charge carried by a Dirac-degenerate BIC (equal to 1/2) is found to be opposite to that of a Dirac-degenerate point, because the interconversion between them is realized by the generation or annihilation of two pairs of *C* points with identical topological charges. The polarization diversity and the topological natures of different types of polarization singularities may open new directions for the study of topological photonic effects in the momentum space, and bring about new opportunities to generate and manipulate vector beams with diverse polarization states and different topological natures.

We are grateful to Professor Shuang Zhang and Dr. Biao Yang for fruitful discussions. The work was supported by National Science Foundation of China (11974428, 61307072, 61405056).

**Figures 1-4:**

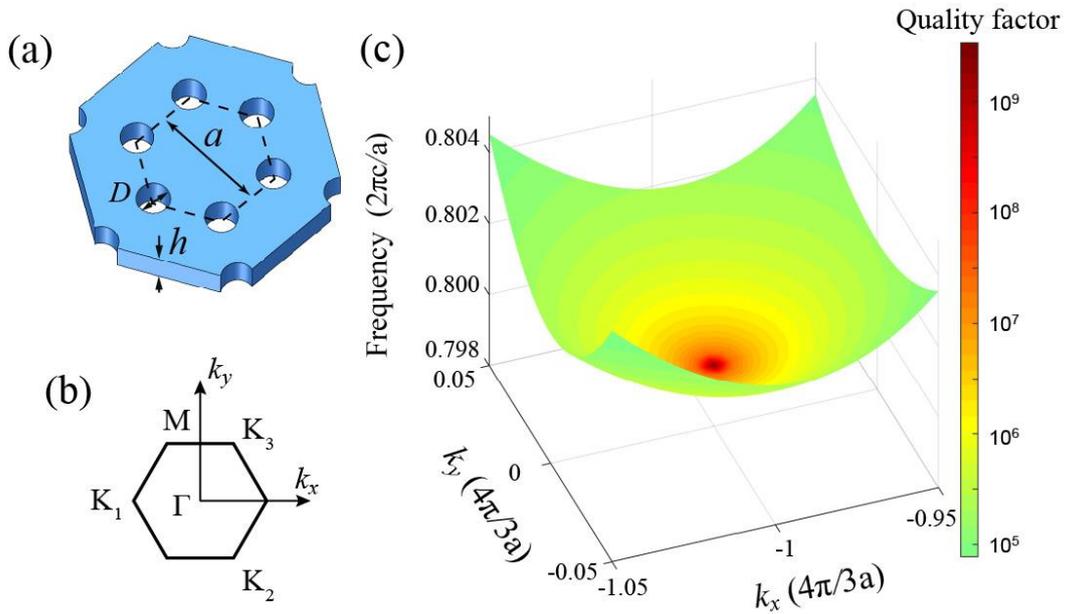

**FIG. 1** Photonic crystal slab (PhCS) supporting a BIC at **K** point. (a) Schematic illustration of the PhCS with a honeycomb array of cylindrical holes etched in a free-standing dielectric slab. (b) The first Brillouin zone in the momentum space of the honeycomb lattice with positions of $\mathbf{K}_1=-K\mathbf{e}_x$, $K=4\pi/(3a)$. (c) Band structure of the PhCSs with ($D$, $h$) equal to ($0.3a$, $0.9074a$) near $\mathbf{K}_1$ point. The color hues represent the quality factors ($Q$ values) of the eigenmodes, which becomes infinite at $\mathbf{K}_1$ point.

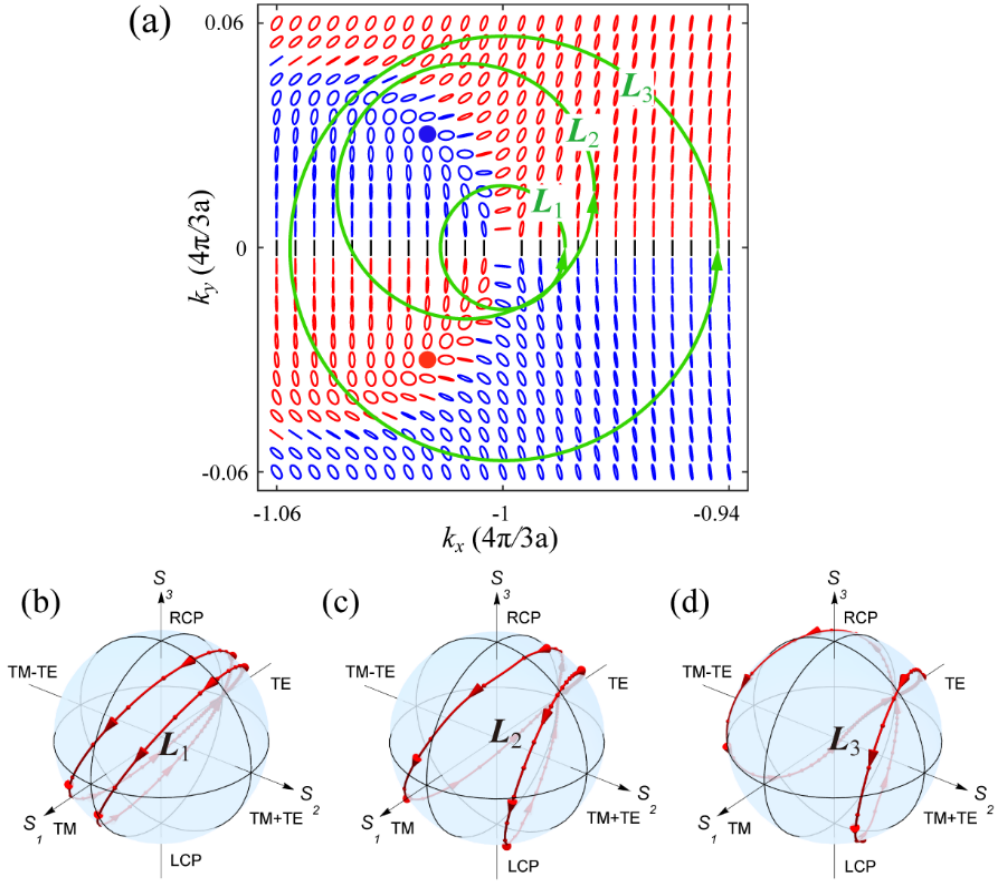

**FIG. 2** Far-field polarization states of the PhCS supporting both BIC and *C* points in the momentum space. (a) Distribution of far-field polarization states of the PhCS near $K_1$-point BIC [Fig. 1(c)]. The red (blue) ellipse denotes the right (left)-handed elliptic polarizations with different orientation angles. The red (blue) disc denotes the right (left)-handed *C* point. (b)-(d) Trajectories of far-field polarization states on the shell of Poincaré sphere when Bloch wave vectors moving along the anticlockwise circles denoted by $L_1$ (b), $L_2$ (c), $L_3$ (d) shown in (a), respectively.

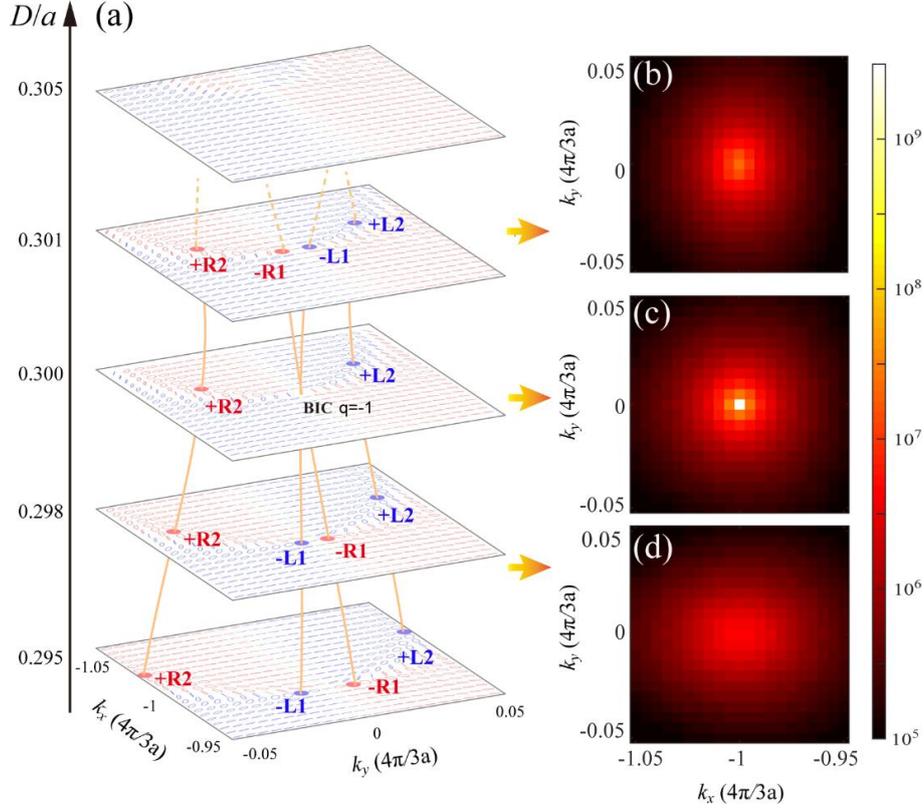

**FIG. 3** Evolutions of *C* points and a BIC on the band of PhCSs near $\mathbf{K}_1$ point with varying diameters *D* and the fixed thickness *h* of 0.9074a (a). Here, red (blue) disc denoted by ±Rn (±Ln) is right- (left-) handed *C* point with a topological charge of ±1/2. Index n is used to denote different pairs of *C* points. The word 'BIC' shows the position of K-point BIC with a topological charge of $q=-1$ in Fig. 2(a). (b),(c),(d) Distributions of *Q* values of eigenmodes supported by the PhCSs with *D* equal to 0.301*a*, 0.3*a* and 0.298*a* near K-point BIC [Fig. 2(a)], respectively.

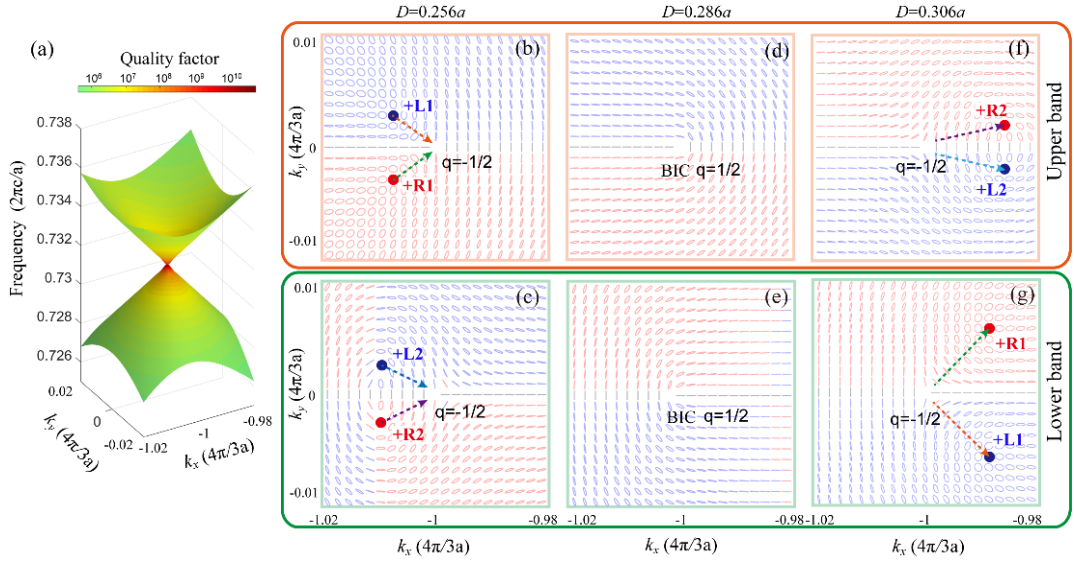

**FIG. 4** Evolutions of *C* points and a double-degenerate BIC on two bands of the PhCS with Dirac cone dispersion. (a) Band structure of the PhCS with (*D*, *h*) equal to (0.286*a*, 1.036*a*) supporting a Dirac-degenerate BIC at **K**$_1$ point. The color hues represent *Q* values of eigenmodes. (b), (d) and (f) Far-field polarization states of eigenmodes in the upper band of PhCSs with the identical *h*=1.036*a* and increasing *D*=0.256*a* (b), 0.286 (d) and 0.306*a* (f), respectively. (c), (e) and (g) Far-field polarization states of eigenmodes in the lower band corresponding to (b), (d) and (f).

# Supplemental Material

## Singular points of polarizations in the momentum space of photonic crystal slabs

Weimin Ye[1*], Yang Gao[4], and Jianlong Liu[2,3†]

1. College of Advanced Interdisciplinary Studies, National University of Defense Technology, Changsha, 410073, China

2. College of Physics, Harbin Institute of Technology, Harbin 150001, China

3. College of Physics and Optoelectronic Engineering, Harbin Engineering University, Harbin 150001, China

4. College of Electronic Engineering, Heilongjiang University, Harbin 150080, China

*wmye72@126.com

†liujl@hit.edu.cn

**Note 1 Band structures of the photonic crystal slab (PhCS) supporting a non-degenerate TE-like odd BIC at $K_1$ point**

Owing to the *z*- and *y*-mirror symmetry (invariance under the operation $\sigma_z$ changing *z* to –*z* and $\sigma_y$ changing *y* to –*y*), all eigenmodes of the honeycomb-lattice PhCS [Fig. 1(a) in the main text] could be divided into TM-like (defined by $\sigma_z=-1$) and TE-like ($\sigma_z=1$) modes. The eigenmodes with Bloch waves along $\Gamma K_1$ direction ($k_y=0$) could be further divided into even (defined by $\sigma_y=1$) and odd ($\sigma_y=-1$) modes. Figure s1(a) and s1(b) show the band structures of TE-like and TM-like modes along $\Gamma K_1$ direction supported by the PhCS with (*D*, *h*) = (0.3*a*, 0.9074*a*). The odd K-point BIC [Fig.1(c)] is located at the bottom of the 3$^{rd}$ band of TE-like modes indicated by a circle [Fig. s1(a)]. Since these leaky modes above the light can couple with incident plane waves from free space, Fano features in the transmittance spectra of the PhCS [Fig. s1(c) and 1(d)] can directly present its band structures. It is worth noting that only s- (p-) polarized plane wave is symmetry-compatible with the odd (even) TE- and TM-like modes of the PhCS.

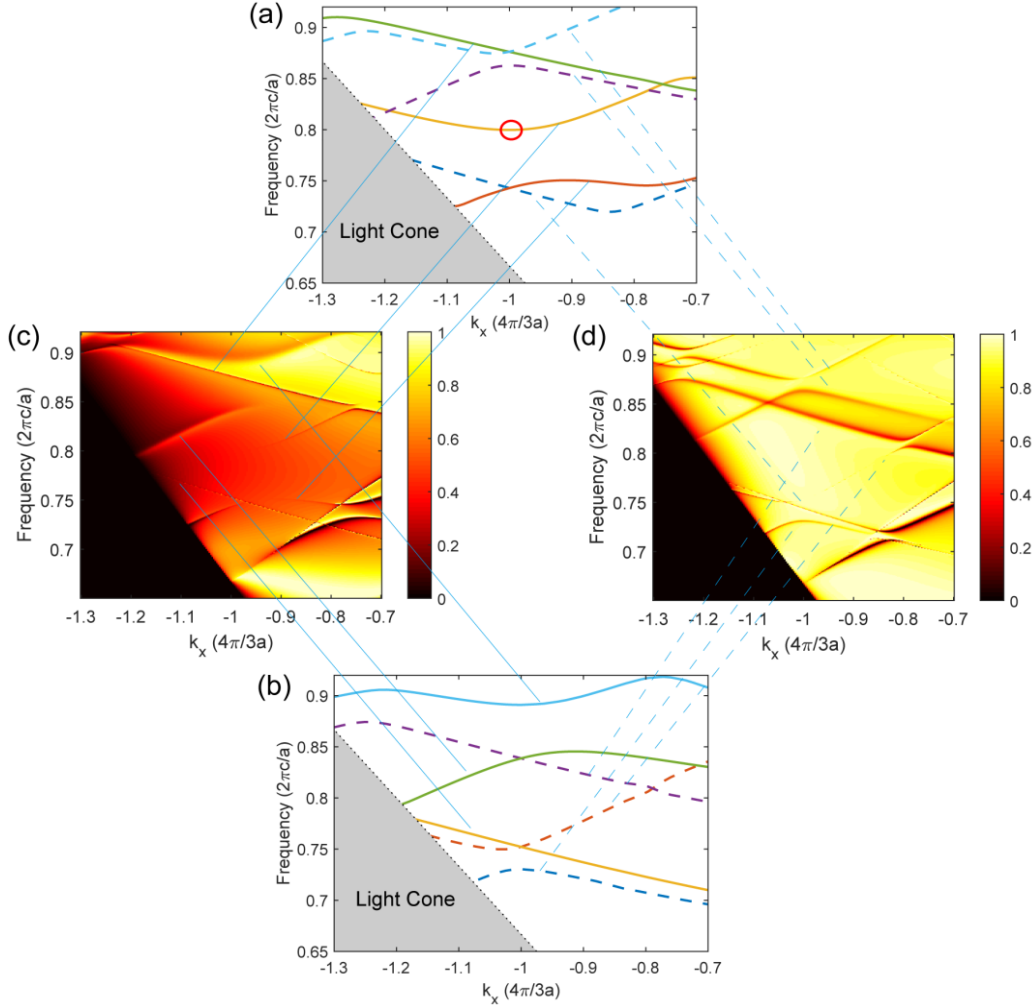

**FIG. s1** Band structures of TE-like (a) and TM-like (b) modes along $\Gamma K_1$ direction supported by the honeycomb-lattice PhCS [Fig. 1(a) in the main text] with $(D, h) = (0.3a, 0.9074a)$. The solid and dash lines are odd and even modes, respectively. (c) and (d) are transmittance spectra of the PhCS, when s-(c) and p-(d) polarized plane waves incident with their in-plane wave vectors along the $\Gamma K_1$ direction, respectively.

**Note 2 Method used to obtain the far-field polarization states**

The far-field polarization states are obtained by the three-dimensional simulations using software COMSOL. The calculation zone in the *x*-*y* plane is a unit cell of the photonic crystal slab, and the Bloch boundary conditions with in-plane wave vector $\boldsymbol{k}(k_x, k_y)$ are used at the boundary of the cell. In the *z* direction, we impose the perfectly matched layers (PML) to absorb the leaking energy of modes. Eigenfreqency solver in COMSOL is used to obtain the eigen field $\boldsymbol{E}(x, y, z)$. Then, the far-field components $\boldsymbol{c}(\boldsymbol{k})$ with the in-plane wave vector $\boldsymbol{k}$ can be obtained by

$$c(k) = (c_x, c_y, c_z) = \frac{1}{\iint_{Cell} dxdy} \iint_{Cell} dxdy e^{-ik_x x - ik_y y} E^*(x, y, z) \tag{s1}$$

The integration is performed over the unit cell on an x-y plane outside the slab. To characterize the polarization state of the far-field components $c(k)$ [Eq. (s1)], we project it into the space spanned by the TE and TM polarized plane wave with identical in-plane wave vector $k$. The projected amplitudes are defined as

$$b^{TE}(k) = e_{TE}(k) \cdot c(k), \quad e_{TE}(k) = \frac{\hat{z} \times k}{|\hat{z} \times k|} = \frac{1}{\sqrt{k_x^2 + k_y^2}} \left(-k_y \hat{x} + k_x \hat{y}\right) \tag{s2a}$$

$$b^{TM}(k) = e_{TM}(k) \cdot c(k), \quad e_{TM}(k) = \frac{k \times e_{TE}(k)}{|k \times e_{TE}(k)|} = \frac{|k|^{-1}}{\sqrt{k_x^2 + k_y^2}} \left[-k_x k_z \hat{x} - k_y k_z \hat{y} + \left(k_x^2 + k_y^2\right) \hat{z}\right] \tag{s2b}$$

Thus, the TE plane wave with wave vector along $\Gamma K_1$ direction is y-polarized.

The Stokes parameters [$S_0$, $S_1$, $S_2$, $S_3$] used to describe the polarization state of the far field with in-plane wave vector $k$ can be obtained from the two projected amplitudes. That is,

$$S_0 = |b^{TE}(k)|^2 + |b^{TM}(k)|^2, \quad S_1 = |b^{TE}(k)|^2 - |b^{TM}(k)|^2 \tag{s3a}$$

$$S_2 = 2\text{Re}\left[b^{TE*}(k) b^{TM}(k)\right], \quad S_3 = 2\text{Im}\left[b^{TE*}(k) b^{TM}(k)\right] \tag{s3b}$$

For the pure polarization state with the orientation angle $\psi$ and ellipticity angle $\chi$, the longitude and latitude of its corresponding position on the shell of Poincaré sphere are equal to $2\psi$ and $2\chi$, respectively. That is,

$$S_0 = 1, \quad S_1 = \cos 2\chi \cos 2\psi, \quad S_2 = \cos 2\chi \sin 2\psi, \quad S_3 = \sin 2\chi \tag{s4}$$

$\psi$ and $\chi$ can be given by the two projected amplitudes[Eqs. (s2a) and (s2b)].

$$2\psi = \phi_- - \phi_+, \quad \tan \chi = \frac{|b_+| - |b_-|}{|b_+| + |b_-|}, \quad b_\pm = |b_\pm| e^{i\phi_\pm} = \frac{1}{\sqrt{2}}\left[b^{TE} \mp i b^{TM}\right] \tag{s5}$$

**Note 3 Realizations of both *C* points and K-point BIC in Fig. 2(a) by the interferences among multiple radiation channels.**

Focusing on the energy conservation and the field inside PhCSs, we utilize a simplified scattering-matrix model (See Ref. [1] which is corresponding to Ref. [24] in the main text) to investigate the realizations of both *C* points and K-point BIC in Fig. 2(a) by the interferences among multiple radiation channels. For a leaky mode of the PhCS with Bloch wave vector $k$, the scattering process at the interface between the slab and the cover (free space) is schematically described by Fig. s2(a). In the cover, only six leakage channels composed of the 0th-order and two 1st-order diffractions are considered. They are the TE and TM polarized plane waves ($|\psi_k^{(TE)}\rangle$ and $|\psi_k^{(TM)}\rangle$) with in-plane wave vectors equal to $k$, $k_1=k+(K_2-K_1)$, $k_2=k+(K_3-K_1)$, respectively. All of them interfere with each other at the interface ($z=h/2$). Inside the PhCS, the field $|v\rangle$ is approximately expressed as a superposition of waveguide modes supported by the identical PhC with an infinity thickness ($h=+\infty$). That is, $|v\rangle = \sum_{m=1}^{N}\left[a_{k,m}^{(+)}e^{i\beta_{k,m}(z-h/2)}|v_{k,m}^{(+)}\rangle + a_{k,m}^{(-)}e^{i\beta_{k,m}(h/2-z)}|v_{k,m}^{(-)}\rangle\right]$. $|v_{k,m}^{(\pm)}\rangle$ denotes the field of the $m^{th}$ waveguide mode propagating along $\pm z$ direction with the real propagation constant $\beta_{k,m}$. $N$ is the amount of the propagating waveguide modes. At the interface ($z=h/2$), these propagating modes partly reflect back into the slab, simultaneously, partly transmit into the cover, which can be described by an interface reflection matrix $r$ ($N\times N$ version) and an interface transmission matrix $t$ ($6\times N$ version). The matrix element $t_{ij}$ quantitatively describes the radiation channel conveying the $j^{th}$ waveguide mode inside the slab into the $i^{th}$ leaky channel in the cover. Taking into account the $z$-mirror symmetry of the slab, the column vector $\boldsymbol{\rho}^{(+)} = \left[a_{k,1}^{(+)}, a_{k,2}^{(+)}, ..., a_{k,N}^{(+)}\right]^T$ composed by the mode coefficients satisfies the eigen equation

$$(\boldsymbol{fr})\boldsymbol{\rho}^{(+)} = \eta\boldsymbol{\rho}^{(+)}, \tag{s6}$$

where the diagonal phase matrix $\boldsymbol{f} = \mathbf{diag}\left[e^{i\beta_{k,1}h}, e^{i\beta_{k,2}h}, ..., e^{i\beta_{k,N}h}\right]$. The eigenvalue $\eta$ is equal to 1 and -1 for TE-like and TM-like modes of the slab, respectively. Meanwhile, the column vector $\boldsymbol{B} = \left[b_k^{(TE)}, b_k^{(TM)}, b_{k_1}^{(TE)}, b_{k_1}^{(TM)}, b_{k_2}^{(TE)}, b_{k_2}^{(TM)}\right]^T$ composed by the mode coefficients of the six leakage channels is given by $\boldsymbol{B} = \boldsymbol{t}\boldsymbol{\rho}^{(+)}$. Thus, the coefficient of every leakage channel is decided by the interference among the $N$ different radiation channels. The total radiating loss of the eigenmode obtained from Eq. (s6) can be characterized by a parameter $Q_R^{-1} = |\boldsymbol{t}\boldsymbol{\rho}^{(+)}|^2 / |\boldsymbol{\rho}^{(+)}|^2$.

For a leaky mode with Bloch wave vector $\boldsymbol{k}(k_x, k_y) = (-1.022K, 0.032K)$ and $\omega a/2\pi c = 0.8015$, the PhCS with $(D, h) = (0.3a, h=+\infty)$ supports twelve propagating waveguide modes. We can obtain the phase matrix $\boldsymbol{f}$ and the interface matrix $\boldsymbol{r}$ ($12\times 12$ version) and $\boldsymbol{t}$ ($6\times 12$ version) of the PhCS by COMSOL simulation (not shown here). Figure s2(b) shows the determinant of the matrix ($\boldsymbol{fr}$-$\boldsymbol{I}$) and the parameter $Q_R$ with the varying thickness $h$ of PhCSs. Based on Eq. (s6), we know that the PhCS with the thickness $h=0.9074a$ supports a TE-like leaky mode with a relatively high quality factor. The calculated ellipticity of its 0th diffraction is equal to -0.96. It is consistent with the existence of a TE-like left-handed $C$ point in Fig. 2(a) in the main text.

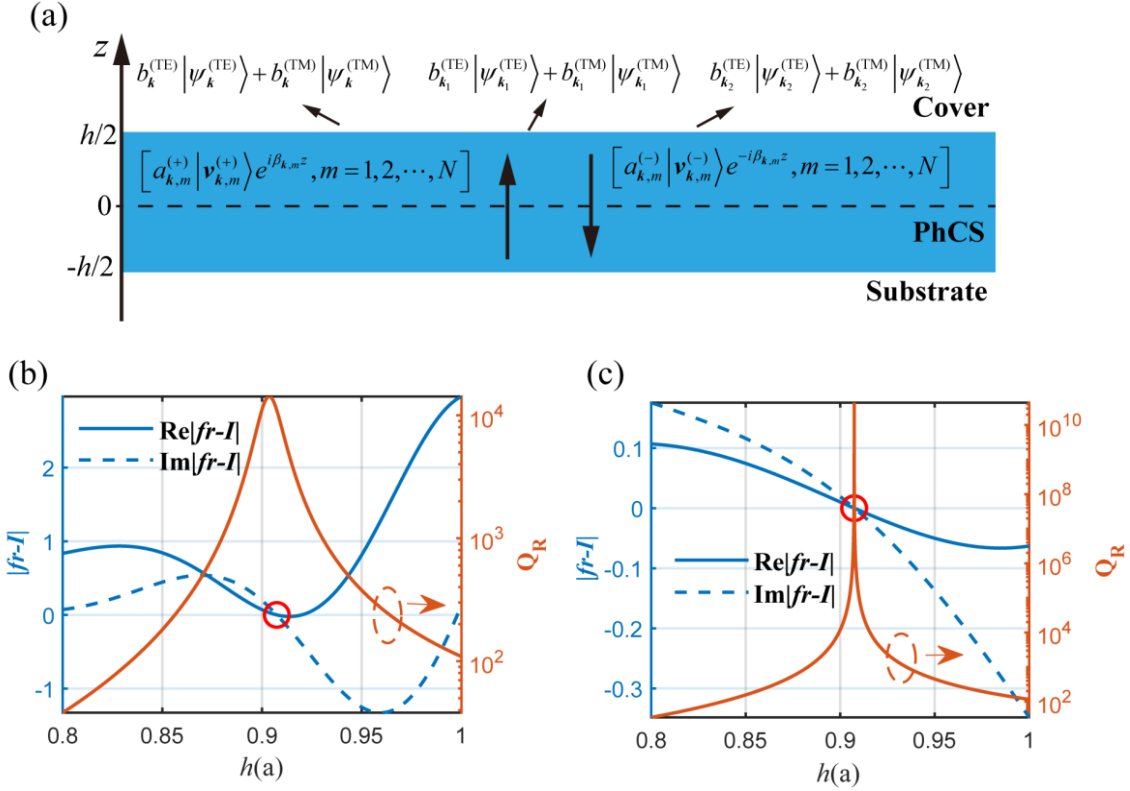

**FIG. s2** Simplified scattering-matrix model of leaky modes with Bloch wave vector $\boldsymbol{k}$ supported by photonic crystals slabs (a). Real part (solid line) and imaginary part (dash line) of the determinant of the matrix ($\boldsymbol{fr}$-**I**) in Eq. (s6) and the parameter $Q_R$ with the varying thickness $h$ of the PhCSs with $D=0.3a$ at $(k_x, k_y)=$ (-1.022$K$, 0.032$K$) with the normalized frequency $\omega a/2\pi c=0.8015$ (b) and at $K_1$ point with $\omega a/2\pi c= 0.7998$ (c), respectively.

The scattering-matrix model can be further simplified by only considering free-space leakages and waveguide modes symmetry-compatible with leaky modes. For a non-degenerate odd mode at $\boldsymbol{K}_1$ point with the normalized frequency $\omega a/2\pi c=0.7998$, only one leakage channel defined by the column vector $\boldsymbol{B} = b_{K_1}^{(TE)}[1,0,1,0,1,0]^T$ and two non-degenerate odd waveguide modes supported by the PhCS with $(D, h)=(0.3a, h=+\infty)$ are symmetry-compatible with it. Thus, the interface reflection matrix $\boldsymbol{r}$ and transmission matrix $\boldsymbol{t}$ are reduced to $2\times 2$ and $1\times 2$ version, respectively. Figure s2(c) presents the determinant of the matrix ($\boldsymbol{fr}$-**I**) and the parameter $Q_R$ with different thickness $h$ of the PhCSs. It agrees with Fig. 1(c) that the PhCS with the thickness $h=0.9074a$ supports a TE-like non-degenerate odd **K**-point BIC, which is achieved by the destructive interferences.

Moreover, at *C* points, the PhCS only couples with the circularly polarized free-space radiations with identical handedness, which can leads to the large difference in reflectance spectra of the PhCS illuminated by the right and left-handed circular polarization (RCP and LCP). Consistent with the existence of a TE-like left-handed *C* point at $\mathbf{k}(k_x, k_y)=(-1.022K, 0.032K)$ and $\omega a/2\pi c=0.8015$ in Fig. 2(a), only the reflectance spectrum of the PhCS under LCP incidence exhibits a Fano line-shape curve at the normalized frequency equal to 0.8017 in Fig. s3.

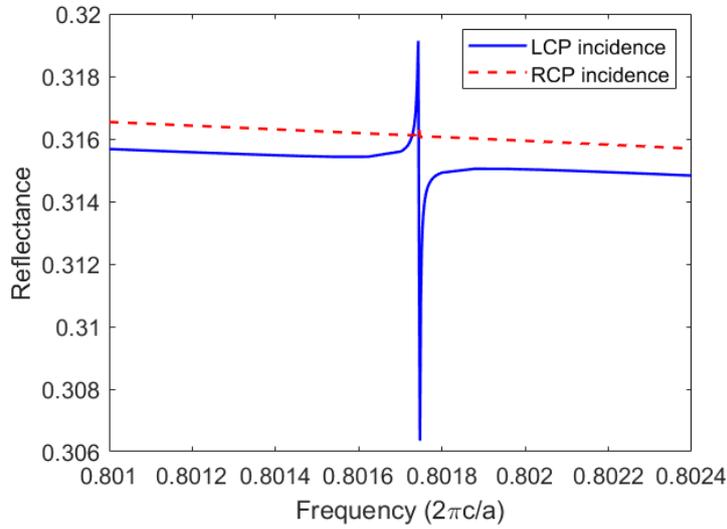

**FIG. s3** Reflectance spectra of the PhCS [Fig. 1(a) in the main text] with $(D, h) = (0.3a, 0.9074a)$, when right- and left-handed circularly polarized (RCP and LCP) plane waves incident with in-plane wave vectors $\mathbf{k}(k_x, k_y)= (-1.022K, 0.032K)$.

**Note 4 Details of Fig. 2(a) in the main text**

For the two-dimensional (2D) photonic crystal slabs (PhCS) with the TE-like non-degenerate odd K-point BICs [Figs. 2(a)] in the main text, figure s4 shows the far-field polarization states of its eigenmodes in the momentum space very close to $\mathbf{K}_1$ point. Here, the dash line denotes the boundary between the right- and left-handedly elliptical polarization states. It is obvious that the far fields of eigenmodes with the Bloch wave vector along the dash line are linear polarizations whose predominant component of electric-field is TM polarization (shown as a short line in the *x* direction). The point of intersection between the dash line and the line with Bloch wave number $k_y=0$ in the momentum space is just $K_1$ point with $(k_x, k_y)$ equal to $(-4\pi/3a, 0)$.

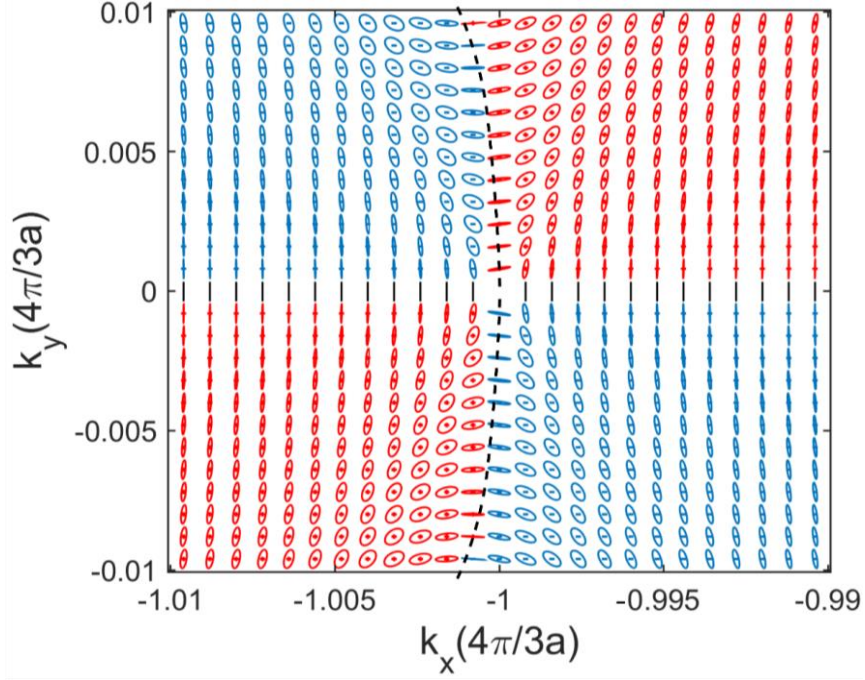

**FIG. s4** Far-field polarization states of eigenmodes very close to $\mathbf{K}_1$ point supported by the PhCS with the identical parameters as in Fig. 2(a) in the main text.

**Note 5 Projected trajectories of three-dimensional (3D) trajectories in Figs. 2(b)-2(d) of the main text on the $S_1$-$S_2$ plane**

By letting $S_3$ equal to zero, we project 3D the trajectories on the shell of the Poincaré sphere to the $S_1$-$S_2$ plane. Here, the winding number of the closed trajectory around the $S_3$ axis is equivalent to the winding number of the projected trajectory around the center of the Poincaré sphere. Figures s5(a)-s5(c) show the projected trajectories of the 3D trajectories shown in Figs. 2(b)-2(d) of the main text, respectively. Because the circle $L_1$ do not enclose any $C$ points, the topological charge $q_B$ of the non-degenerate odd BIC is equal to half of the winding number of the trajectory [Fig. 2(b) and Fig. s5(a)]. Note that the projected trajectory is clockwise. Thus, $q_B$=-1. Both Fig. 2(c) and Fig. s5(b) show the winding number of the trajectory corresponding to the circle $L_2$ enclosing the BIC and one C point is -1. It means the topological charge $q_C$ of $C$ points is 1/2. Therefore, the winding number of the trajectory [Fig. 2(d) and Fig. s5(c)] corresponding to the circle $L_3$ enclosing the BIC and two C points must be zero.

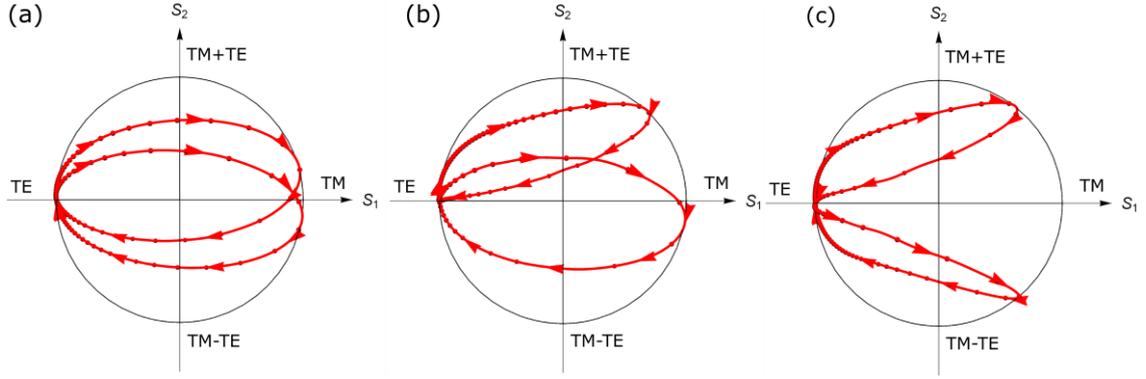

**FIG. s5** Projected trajectories on the $S_1-S_2$ plane. (a)-(c) Corresponding to the three-dimensional trajectories shown in Figs. 2(b)-2(d) in the main text, respectively.

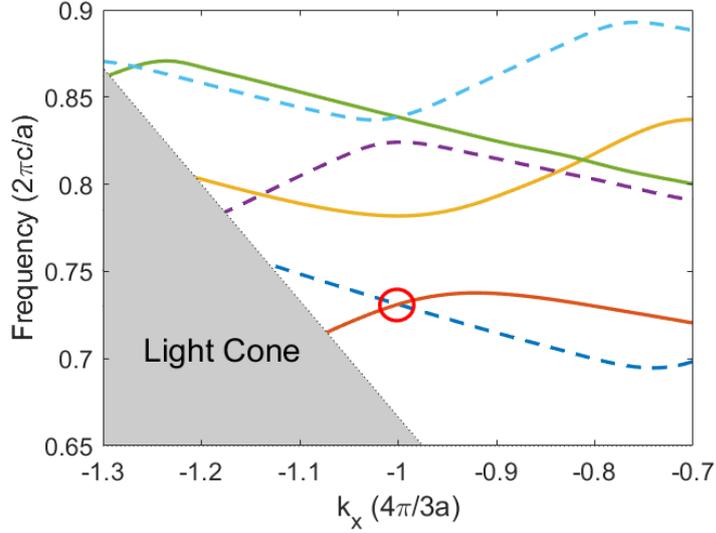

**FIG. s6** Band structures of TE-like modes supported by the honeycomb-lattice PhCS [in Fig. 1(a)] with ($D$, $h$) equal to (0.286$a$, 1.036$a$). The solid and dash lines are odd and even modes, respectively. A TE-like Dirac-degenerate BIC with a normalized frequency of 0.7311 emerges at $\mathbf{K}_1$ point [Fig. 4(a) in the main text], which is indicated by a circle.

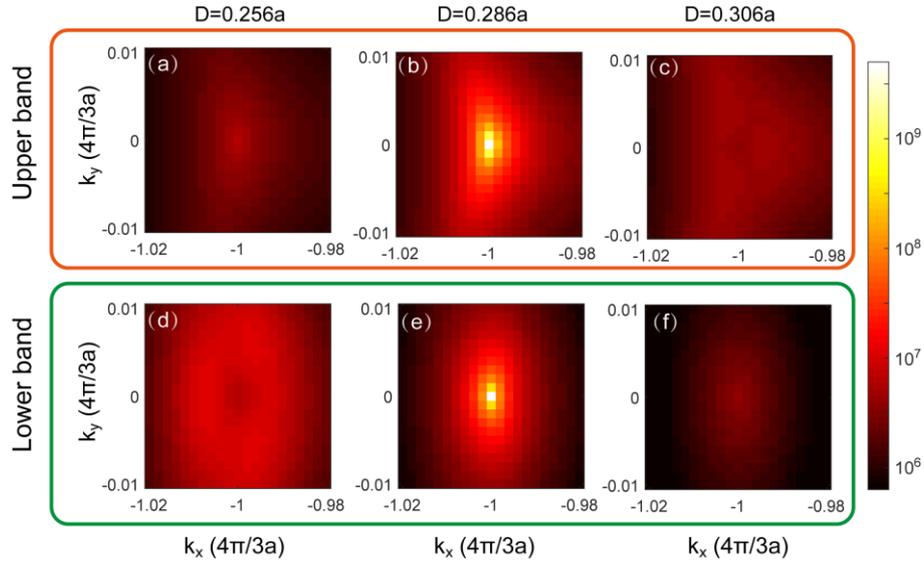

**FIG. s7** Distributions of quality factors (*Q* values) of eigenmodes on the upper (a-c) and lower (d-f) band of the PhCSs with Dirac cone dispersion [Fig. 4 in the main text]. The Dirac-degenerate BIC with infinite *Q* value appears when *D* equals to 0.286.